\documentclass[twocolumn,preprint,3p,twocolumn]{elsarticle}
\usepackage[latin9]{inputenc}
\usepackage{amsmath}
\usepackage{amssymb}
\usepackage{graphicx}
\usepackage{esint}

\makeatletter

\usepackage{color}\usepackage{epsfig}\usepackage{dcolumn}\usepackage{bm}\usepackage{color}

\makeatother

\begin{document}
\begin{frontmatter}

\title{Effect of Interdots Electronic Repulsion in the Majorana Signature
for a Double Dot Interferometer}

\author{L. S. Ricco$^{1}$, Y. Marques$^{1}$, F. A. Dessotti$^{1}$, M.
de Souza$^{2,*}$ and A. C. Seridonio\corref{A}$^{1,2}$}

\address{$^{1}$Departamento de F\'{i}sica e Qu\'{i}mica, Unesp - Univ Estadual
Paulista, 15385-000, Ilha Solteira, SP, Brazil\\
 $^{2}$IGCE, Unesp - Univ Estadual Paulista, Departamento de F\'{i}sica,
13506-900, Rio Claro, SP, Brazil\\
 }

\cortext[A]{ Current address: Institute of Semiconductor and Solid State Physics,
Johannes Kepler University Linz, Austria.}

\cortext[A]{Corresponding Author: seridonio@dfq.feis.unesp.br}
\begin{abstract}
We investigate theoretically the features of the Majorana hallmark
in the presence of Coulomb repulsion between two quantum dots describing
a spinless double dot interferometer, where one of the dots
is strongly coupled to a Kitaev wire within the topological phase.
Such a system has been originally proposed without Coulomb interaction
in J. of Appl. Phys.\textbf{ 116}, 173701 (2014). Our findings reveal
that for dots in resonance, the ratio between the strength of Coulomb
repulsion and the dot-wire coupling changes the width of the Majorana
zero-bias peak for both Fano regimes studied, indicating thus that
the electronic interdots correlation influences the
Majorana state lifetime in the dot hybridized with the wire. Moreover, for
the off-resonance case, the swap between the energy levels of the
dots also modifies the width of the Majorana peak, which does not
happen for the noninteracting case. The results obtained here can
guide experimentalists that pursuit a way of revealing Majorana signatures.\end{abstract}
\begin{keyword}
Kitaev wire,  double dot interferometer, Majorana bound states,
quantum dots, Fano effect \PACS 85.35.Be, 74.78.-w, 85.25.Dq, 73.23.Hk,
73.63.Kv
\end{keyword}
\end{frontmatter}

\section{Introduction}

Recently, the pursuit for Majorana quasiparticles in condensed matter
systems has attracted a lot of attention, since they are one of the
most promising candidate to build a quantum bit, the fundamental structure
of quantum computation \citep{key-1,key-2}. In this sense, topological
superconductors have been broadly studied, once they can host zero-energy
modes of Majorana bound states (MBSs) in their edges {[}3-5{]}. The
Kitaev wire within the topological phase is an example \citep{key-3},
since in such a proposal a 1D topological \emph{p}-wave superconductor
gives rise to MBSs attached to its edges. Experimentally, this setup
can be implemented by putting a semiconductor nanowire, with strong
spin-orbit coupling, close to an \emph{s}-wave superconductor and
under an external magnetic field. In this situation, \emph{p}-wave
topological superconductivity is induced in the nanowire by the so-called
proximity effect {[}6-9{]}.

The MBSs can be detected in electronic transport measurements by attaching
quantum objects to them, as for instance quantum dots (QDs) {[}10-21{]}.
In the case of a single QD coupled to a MBS, a zero-bias anomaly (ZBA)
is theoretically predicted to appear in the conductance, with amplitude
$0.5G_{0}$, where $G_{0}=e^{2}/h$ is the quantum of conductance
\citep{key-13}. The emergence of the ZBA is due to the leaking of
the MBS zero-mode into the QD \citep{key-14}. Such an anomaly was
first detected experimentally by Mourik \emph{et al.} \citep{key-19},
where the MBSs are supposed to exist once the ZBA in the conductance
persists even at high gate voltages and magnetic fields. However,
another physical phenomena can lead to the ZBA, as for instance the
Kondo effect. Within this perspective, the detection of Majorana excitations
becomes inconclusive.

Alternatives have been proposed in order to obtain the MBSs, involving
ferromagnetic chains on top of \emph{s}-wave superconductors with
strong spin-orbit parameter {[}23-27{]}. Such a setup allows the emulation
of the Kitaev wire, once the \emph{p}-wave topological superconductivity
is induced in the ferromagnetic atoms, yielding MBSs at the chain
boundaries. Recently, this proposal has been accomplished experimentally,
by using Fe atoms on a Pb superconductor surface \citep{key-26}.
The electronic conductance features have been measured by using an
STM tip, in which the ZBA was verified at the ends of the Fe chain,
thus suggesting the presence of MBSs. However, the ZBA obtained cannot
be associated exclusively to the presence of MBSs \citep{key-28}
and, therefore the issue of Majorana detection has not been solved
completely yet.

\begin{figure}[!htb]
\begin{centering}
\includegraphics[height=5cm]{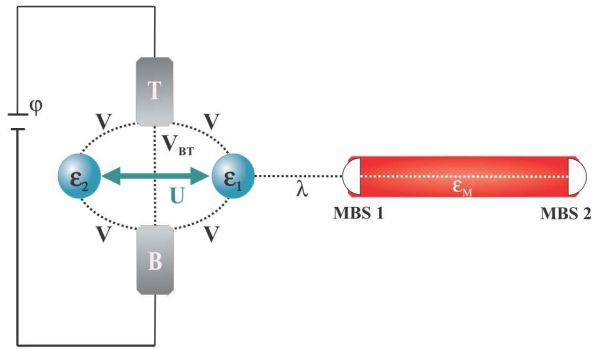} \protect\protect\caption{\label{fig1}(Color online) Sketch of the system proposed: a Kitaev
wire hosting Majorana bound states (MBSs) in its edges (white half-spheres)
side-coupled to a spinless double dot interferometer. This
device is composed by metallic leads (Bottom(B) and Top(T)) and two
quantum dots, with energy levels $\varepsilon_{1}$ and $\varepsilon_{2}$,
respectively. $U$ represents the intensity of the interdots Coulomb
repulsion, $V$ is the tunneling amplitude between the QDs and leads.
The coupling lead-lead is given by $V_{BT}.$ The wire is coupled
to the QD 1 by the strength $\lambda.$ The overlap between the wave
functions of the MBS 1 and MBS 2 is denoted by $\varepsilon_{M}$
and $\varphi$ is the bias-voltage of the setup. }

\par\end{centering}

\centering{}
\end{figure}

In this scenario, in order to detect MBSs new proposals become necessary.
In our previous work \citep{key-16}, we proposed theoretically a
helpful tool to detect MBSs in a system composed by a spinless double dot
interferometer, with two noninteracting QDs, where one of them is
coupled to one edge of the Kitaev wire within the topological phase.
Such a spinless system can be achieved experimentally by applying
a magnetic field strong enough to provide a Zeeman splitting in the
energy levels of the QDs and metallic leads as well. Thus, just one
spin orientation prevails. We found that the ZBA is robust and independent
on the Fano regime of interference \citep{key-29}. Additionally,
we developed a novel manner to verify the presence of MBSs, which
looks beyond the ZBA signature: through simulations of transmittance
as a function of the detuning for the energy levels of the QDs and
Fermi energy for the metallic leads, we found that a MBS has a particular
way of breaking the symmetry of such transmittance profiles, which
can be experimentally accessed by conductance measurements. Here we
explore the same device studied in the previous work (Fig.\,\ref{fig1}),
but now we consider the Coulomb repulsion between the QDs employing
the Hubbard I approximation \citep{key-30} in order to close the
system of Green functions. Such a mean field approach is valid for
$T\gg T_{K}$, where $T_{K}$ is the Kondo temperature \citep{KONDO}. Interestingly enough, it is possible to observe a Kondo peak
even in absence of the spin degree of freedom in our double dot system, i.e, in the spinless regime. Such a specific condition is requested to ensure the topological superconductivity in the framework of the Kitaev wire. According to Ref.\citep{KONDO}, the spinless Hamiltonian for a double dot setup can be mapped into an effective one describing a single dot, in which the spin degree of freedom is restored, thus allowing the possibility of the Kondo effect when $T\ll T_{K}.$  In this case, the \textit{pseudo} localized moment of the dot is screened by those from the effective conduction band of the model via an antiferromagnetic exchange interaction, as in the standard Kondo effect. Such a phenomenon can be prevented for $T\gg T_{K},$ which is the range wherein this work focus on, thus ensuring the employment of the Hubbard I approach. Such a method is a well-established procedure in the literature that allows to catch the key features concerning the Physics away from the Kondo limit. Otherwise, the ZBA would be the outcome of a competition between the Majorana and Kondo phenomena \citep{U-Vernek}.

Our findings reveal that the ZBA $0.5$ characteristic amplitude of
the MBS remains even in the presence of Coulomb repulsion between
the QDs, with slight fluctuations around such a value. However, for
the case of QDs in resonance, the ratio between the Coulomb repulsion
and the QD-wire coupling modifies the ZBA width, revealing thus that
the electronic interdots repulsion affects the Majorana state lifetime
in the QD. Moreover, in the interacting system the width of the Majorana
signature also is influenced by swapping the energy levels of the
QDs, which does not occur without Coulomb repulsion.

\section{Theoretical Model}

To describe the system presented in Fig.\,\ref{fig1} we use the Hamiltonian
inspired on the original proposal from Liu and Baranger \citep{key-13},

\begin{align}
\mathcal{H} & =\sum_{\alpha k}\tilde{\varepsilon}_{\alpha k}c_{\alpha k}^{\dagger}c_{\alpha k}+\sum_{j}\varepsilon_{j}d_{j}^{\dagger}d_{j}+Ud_{1}^{\dagger}d_{1}d_{2}^{\dagger}d_{2}\nonumber \\
 & +\mathcal{H}_{\text{{lead-dot}}}+\mathcal{H}_{\text{{lead-lead}}}+\mathcal{H}_{\text{{MBSs}}},\label{eq:TIAM}
\end{align}
where the operator $c_{\alpha k}^{\dagger}$ ($c_{\alpha k}$) creates
(annihilates) an electron in the lead $\alpha$= B/T (Bottom/Top),
with energy $\tilde{\varepsilon}_{\alpha k}=\varepsilon_{k}-\mu_{\alpha}$,
wherein $\mu_{\alpha}$ as the chemical potential and $k$ is the
wave number. We consider $\mu_{B}-\mu_{T}=2\Delta\mu=e\varphi$ as
the bias between the leads, where $e>0$ is the electron charge and
$\varphi$ is the bias-voltage. $d_{j}^{\dagger}$ ($d_{j}$) creates
(annihilates) an electron in the state $\varepsilon_{j}$ in the QDs
and $U$ is the interdots Coulomb repulsion, with $j=1,2.$ $\mathcal{H}_{\text{{lead-dot}}}=V\sum_{\alpha kj}(c_{\alpha k}^{\dagger}d_{j}+\text{{H.c.}}),$
where $V$ is the coupling amplitude between the leads and the QDs,
and $\mathcal{H}_{\text{{lead-lead}}}=V_{BT}\sum_{kp}(c_{Bk}^{\dagger}c_{Tp}+\text{{H.c.}})$,
with $V_{BT}$ the direct lead-lead hybridization. Furthermore,

\begin{equation}
\mathcal{H}_{\text{{MBSs}}}=i\varepsilon_{M}\Psi_{1}\Psi_{2}+\lambda(d_{1}-d_{1}^{\dagger})\Psi_{1}\label{eq:MBSs}
\end{equation}
describes an effective model for the Kitaev wire within the topological
phase, where $\Psi_{l}=\Psi_{l}^{\dagger}$ is the Majorana operator,
with $l=1,2.$ The MBS 1 $\Psi_{1}$ and MBS 2 $\Psi_{2}$ are connected
via $\varepsilon_{M}\sim e^{-L/\xi}$, where $L$ is the distance
between them and $\xi$ is the superconductor coherence length. The
coupling strength between the MBS 1 and QD 1 is $\lambda.$ According
to the Landauer-Büttiker formula \citep{key-31} for the zero-bias
regime, the conductance depends on the transmittance $\mathcal{T}(\varepsilon)$
as follows:

\begin{equation}
G=G_{0}\int d\varepsilon\left(-\frac{\partial f_{F}}{\partial\varepsilon}\right)\mathcal{T}(\varepsilon),\label{eq:CONDUCTANCE}
\end{equation}
where $f_{F}$ is the Fermi-Dirac distribution.

In order to obtain the transmittance, within the wide band limit,
we perform the transformations $c_{Bk}=\frac{1}{\sqrt{2}}(c_{ek}+c_{ok})$
and $c_{Tk}=\frac{1}{\sqrt{2}}(c_{ek}-c_{ok})$ in the Hamiltonian
of Eq.\,(\ref{eq:TIAM}), which depends on the \textit{even} and \textit{odd}
conduction operators $c_{ek}$ and $c_{ok}$, respectively. Thus,
Eq.\,(\ref{eq:TIAM}) becomes $\mathcal{H}=\mathcal{H}_{e}+\mathcal{H}_{o}+\mathcal{\tilde{H}}_{\text{{tun}}}=\mathcal{H}_{\varphi=0}+\mathcal{\tilde{H}}_{\text{{tun}}},$
where

\begin{align}
\mathcal{H}_{e}=\sum_{k}\varepsilon_{k}c_{ek}^{\dagger}c_{ek}+\sum_{j}\varepsilon_{j}d_{j}^{\dagger}d_{j}+ & Ud_{1}^{\dagger}d_{1}d_{2}^{\dagger}d_{2}\nonumber \\
+\sqrt{2}V\sum_{jk}(c_{ek}^{\dagger}d_{j}+\text{{H.c.}})+V_{BT}\sum_{kp}c_{ek}^{\dagger}c_{ep} & +\mathcal{H}_{\text{{MBSs}}}\nonumber \\
\label{eq:even}
\end{align}
describes effectively the couplings between leads and QDs via the
strength $\sqrt{2}V$ and $\mathcal{H}_{o}=\sum_{k}\varepsilon_{k}c_{ok}^{\dagger}c_{ok}-V_{BT}\sum_{kp}c_{ok}^{\dagger}c_{op}$
is for the part of the system decoupled from the QDs. $\mathcal{H}_{e}$
is connected to $\mathcal{H}_{o}$ by the tunneling Hamiltonian $\mathcal{\tilde{H}}_{\text{{tun}}}=-\Delta\mu\sum_{k}(c_{ek}^{\dagger}c_{ok}+c_{ok}^{\dagger}c_{ek}),$
which in the zero-bias regime is perturbative, since $\Delta\mu\rightarrow0$
due $\varphi\rightarrow0.$ Thus, according to the linear response
theory and by applying the equation of motion method (EOM) \citep{key-31},
the transmittance \citep{key-16} is given by

\begin{align}
\frac{\mathcal{T}\left(\varepsilon\right)}{\mathcal{T}_{b}} & =1+(1-q_{b}^{2})\tilde{\Gamma}\sum_{j\tilde{j}}\text{{Im}}(\tilde{\mathcal{G}}_{d_{j},d_{\tilde{j}}})\nonumber \\
 & +2q_{b}\tilde{\Gamma}\sum_{j\tilde{j}}\text{{Re}}(\tilde{\mathcal{G}}_{d_{j},d_{\tilde{j}}}),\label{eq:TRANSMITTANCE}
\end{align}
where $\mathcal{T}_{b}=\frac{4x}{\left(1+x\right)^{2}}$ represents
the background transmittance with $x=(\pi\rho_{0}V_{BT})^{2}$, $\rho_{0}$
is the leads density of states, $\tilde{\Gamma}=\frac{\Gamma}{1+x}$
is an effective QD-leads coupling, with $\Gamma=2V^{2}\pi\rho_{0}$,
and $q_{b}=\sqrt{\frac{\mathcal{R}_{b}}{\mathcal{T}_{b}}}=\frac{\left(1-x\right)}{2\sqrt{x}}$
is the Fano parameter \citep{key-29,key-32}, where $\mathcal{R}_{b}$
represents the corresponding background reflectance.

With the aim to get the retarded Green functions of the system, we
first express the Majorana operators $\Psi_{1}$ and $\Psi_{2}$ in
terms of a nonlocal regular fermion state $f$, according to the following
relations: $\Psi_{1}=\frac{1}{\sqrt{2}}(f^{\dagger}+f)$ and $\Psi_{2}=i\frac{1}{\sqrt{2}}(f^{\dagger}-f),$
with $f\neq f^{\dagger}$ and $[f,f^{\dagger}]_{+}=1.$ Then, Eq.\,(\ref{eq:MBSs})
becomes $\mathcal{H}_{\text{{MBSs}}}=\varepsilon_{M}(f^{\dagger}f-\frac{1}{2})+\frac{\lambda}{\sqrt{2}}(d_{1}f^{\dagger}+fd_{1}^{\dagger})+\frac{\lambda}{\sqrt{2}}(d_{1}f-d_{1}^{\dagger}f^{\dagger}).$

The EOM procedure \citep{key-31} can be summarized as

\begin{equation}
(\varepsilon+i0^{+})\tilde{\mathcal{G}}_{\mathcal{AB}}=[\mathcal{A},\mathcal{B^{\dagger}}]_{+}+\tilde{\mathcal{G}}_{\left[\mathcal{A},\mathcal{\mathcal{H}}_{i}\right]\mathcal{B}}\label{eq:EOM}
\end{equation}
where $\tilde{\mathcal{G}}_{\mathcal{AB}}$ represents the retarded
Green function in the energy domain $\varepsilon$, with $\mathcal{A}$
and $\mathcal{B}$ as fermionic operators belonging to the Hamiltonian
$\mathcal{\mathcal{H}}_{i}$. The Green function for the QD in the
time domain $t$ is definite by

\begin{equation}
\mathcal{G}_{d_{j}d_{l}}(t)=-\frac{i}{\hbar}\theta\left(t\right){\tt Tr}\{\varrho_{\text{{e}}}[d_{j}\left(t\right),d_{l}^{\dagger}\left(0\right)]_{+}\},\label{eq:Gdjdj}
\end{equation}
wherein $\theta\left(t\right)$ is the Heaviside step function and
$\varrho_{\text{e}}$ is the density-matrix for Eq.\,(\ref{eq:even}).
By applying the EOM procedure in Eq.\,(\ref{eq:Gdjdj}), we obtain

\begin{align}
(\varepsilon-\varepsilon_{j}-\Sigma-\delta_{j1}\Sigma_{\text{{MBS1}}})\tilde{\mathcal{G}}_{d_{j}d_{l}}=\delta_{jl}+ & \Sigma[\underset{\tilde{l}\neq j}{\sum}\tilde{\mathcal{G}}_{d_{j}d_{\tilde{l}}}]\nonumber \\
+U[\tilde{\mathcal{G}}_{d_{j}n_{\bar{j}},d_{l}}+\lambda^{2}\tilde{K}\delta_{j1}\tilde{\mathcal{G}}_{d_{j}^{\dagger}n_{\bar{j}},d_{l}}]\label{eq:Gjl}
\end{align}
where the index $\bar{j}$ represents the opposite of $j$, i.e, $j\leftrightarrow\bar{j}\equiv1\leftrightarrow2$,
with $\Sigma=-\frac{(\sqrt{x}+i)}{1+x}\Gamma$ and $\Sigma_{\text{{MBS1}}}=\lambda^{2}K(1+\lambda^{2}\tilde{K})$
as the self-energies that appear in the noninteracting case \citep{key-16},
where $K=\frac{1}{2}\left(\frac{1}{\varepsilon-\varepsilon_{M}+i0^{+}}+\frac{1}{\varepsilon+\varepsilon_{M}+i0^{+}}\right)$,
$\tilde{K}=\frac{K}{\varepsilon+\varepsilon_{1}+\tilde{\Sigma}-\lambda^{2}K}$
and $\tilde{\Sigma}$ is the complex conjugate of $\Sigma.$ We point
out that making $U=0$ in Eq.\,(\ref{eq:Gjl}), we obtain the same
expression of the noninteracting system {[}Eq.(17) of Ref.\,\cite{key-16}{]}.
By applying the EOM approach in the same way we have performed above,
we obtain the retarded Green functions of four operators $\tilde{\mathcal{G}}_{d_{j}n_{\bar{j}},d_{l}}$
and $\tilde{\mathcal{G}}_{d_{j}^{\dagger}n_{\bar{j}},d_{l}}$, given
by

\begin{align}
(\varepsilon-\varepsilon_{j}-U+i0^{+})\tilde{\mathcal{G}}_{d_{j}n_{\bar{j}},d_{l}} & =\delta_{jl}\bigl\langle n_{\bar{j}}\bigr\rangle\nonumber \\
+\sqrt{2}V\sum_{k}[\tilde{\mathcal{G}}_{n_{\bar{j}}c_{ek},d_{l}}+\tilde{\mathcal{G}}_{d_{\bar{j}}^{\dagger}c_{ek}d_{j},d_{l}} & -\tilde{\mathcal{G}}_{c_{ek}^{\dagger}d_{\bar{j}}d_{j},d_{l}}]\nonumber \\
+\frac{\lambda}{\sqrt{2}}[\delta_{\bar{j}1}\tilde{\mathcal{G}}_{f^{\dagger}d_{\bar{j}}d_{j},d_{l}}-\delta_{\bar{j}1}\tilde{\mathcal{G}}_{d_{\bar{j}}^{\dagger}fd_{j},d_{l}} & -\delta_{j1}\tilde{\mathcal{G}}_{n_{\bar{j}}f,d_{l}}]\nonumber \\
+\frac{\lambda}{\sqrt{2}}[\delta_{\bar{j}1}\tilde{\mathcal{G}}_{\eta d_{\bar{j}}d_{j},d_{l}}-\delta_{\bar{j}1}\tilde{\mathcal{G}}_{d_{\bar{j}}^{\dagger}\eta^{\dagger}d_{j},d_{l}} & -\delta_{j1}\tilde{\mathcal{G}}_{n_{\bar{j}}\eta^{\dagger},d_{l}}]\nonumber \\
\label{eq:FOUR_1}
\end{align}
and

\begin{align}
(\varepsilon+\varepsilon_{j}+U+i0^{+})\tilde{\mathcal{G}}_{d_{j}^{\dagger}n_{\bar{j}},d_{l}}= & \delta_{\bar{j}l}\bigl\langle d_{j}^{\dagger}d_{\bar{j}}^{\dagger}\bigr\rangle\nonumber \\
+\sqrt{2}V\sum_{k}[\tilde{\mathcal{G}}_{d_{\bar{j}}^{\dagger}c_{ek}d_{j}^{\dagger},d_{l}}-\tilde{\mathcal{G}}_{c_{ek}^{\dagger}d_{\bar{j}}d_{j}^{\dagger},d_{l}} & -\tilde{\mathcal{G}}_{n_{\bar{j}}c_{ek}^{\dagger},d_{l}}]\nonumber \\
+\frac{\lambda}{\sqrt{2}}[\delta_{j1}\tilde{\mathcal{G}}_{n_{\bar{j}}f^{\dagger},d_{l}}-\delta_{\bar{j}1}\tilde{\mathcal{G}}_{d_{\bar{j}}f^{\dagger}d_{j}^{\dagger},d_{l}}+ & \delta_{\bar{j}1}\tilde{\mathcal{G}}_{fd_{\bar{j}}^{\dagger}d_{j}^{\dagger},d_{l}}]\nonumber \\
+\frac{\lambda}{\sqrt{2}}[\delta_{j1}\tilde{\mathcal{G}}_{n_{\bar{j}}f,d_{l}}-\delta_{\bar{j}1}\tilde{\mathcal{G}}_{d_{\bar{j}}fd_{j}^{\dagger},d_{l}}+ & \delta_{\bar{j}1}\tilde{\mathcal{G}}_{f^{\dagger}d_{\bar{j}}^{\dagger}d_{j}^{\dagger},d_{l}}].\nonumber \\
\label{eq:FOUR_2}
\end{align}
with expectation values

\begin{equation}
\bigl\langle n_{\bar{j}}\bigr\rangle=\bigl\langle d_{\bar{j}}^{\dagger}d_{\bar{j}}\bigr\rangle=(-\frac{1}{\pi})\int_{-\infty}^{\infty}\text{{d\ensuremath{\varepsilon}}}f_{F}(\varepsilon)\text{{Im}}\{\tilde{\mathcal{G}}_{d_{\bar{j}}d_{\bar{j}}}\}\label{eq:nj_bar}
\end{equation}
and

\begin{equation}
\bigl\langle d_{j}^{\dagger}d_{\bar{j}}^{\dagger}\bigr\rangle=(-\frac{1}{\pi})\int_{-\infty}^{\infty}\text{{d\ensuremath{\varepsilon}}}f_{F}(\varepsilon)\text{{Im}}\{\tilde{\mathcal{G}}_{d_{j}^{\dagger}d_{\bar{j}}}\}.\label{eq:Cooper:average}
\end{equation}

As one can see in Eqs.\,(\ref{eq:nj_bar}) and (\ref{eq:Cooper:average}),
a self-consistent calculation is required to obtain the electronic
occupation number for the QDs $\bigl\langle n_{\bar{j}}\bigr\rangle=\bigl\langle d_{\bar{j}}^{\dagger}d_{\bar{j}}\bigr\rangle$
and the expectation value of the delocalized Cooper paring $\bigl\langle d_{j}^{\dagger}d_{\bar{j}}^{\dagger}\bigr\rangle$.
As the latter amount is induced in the QDs due to the presence of
the Kitaev wire, such a quantity is expected to be smaller compared
to the usual occupation number \citep{U-Vernek}. We have confirmed
this feature in the self-consistent process.

In order to close the system of Green functions, we apply an approximation
method motivated by the Hubbard I decoupling \citep{key-30}, which
can be summarized as $\tilde{\mathcal{G}}_{\mathcal{ABCD}}=\left\langle \mathcal{AB}\right\rangle \tilde{\mathcal{G}}_{\mathcal{CD}}$,
with the property $\left\langle \mathcal{AB}\right\rangle =\left\langle \mathcal{B}^{\dagger}\mathcal{A}^{\dagger}\right\rangle $.
After this procedure, we obtain:

\begin{align}
(\varepsilon-\varepsilon_{j}-\Sigma^{(j)}-\delta_{j1}\Sigma_{\text{{MBS1}}}^{\text{{full}}})\tilde{\mathcal{G}}_{d_{j}d_{j}}= & 1\nonumber \\
+\frac{U\bigl\langle n_{\bar{j}}\bigr\rangle}{(\varepsilon-\varepsilon_{j}-U+i0^{+})}+\Sigma^{(j)}\tilde{\mathcal{G}}_{d_{j}d_{\bar{j}}}\label{eq:Gjj}
\end{align}
and

\begin{align}
(\varepsilon-\varepsilon_{j}-\Sigma^{(j)}-\delta_{j1}\Sigma_{\text{{MBS1}}}^{\text{{full}}})\tilde{\mathcal{G}}_{d_{j}d_{\bar{j}}}=\nonumber \\
\delta_{j1}U\lambda^{2}\tilde{K}_{U}\bigl\langle d_{j}^{\dagger}d_{\bar{j}}^{\dagger}\bigr\rangle(\frac{U\bigl\langle n_{\bar{j}}\bigr\rangle}{\varepsilon-\varepsilon_{j}-U+i0^{+}}+1)\nonumber \\
+\Sigma^{(j)}\tilde{\mathcal{G}}_{d_{\bar{j}}d_{\bar{j}}},\label{eq:Gjjbar}
\end{align}
where $\tilde{K}_{U}=\frac{\tilde{K}}{(\varepsilon+\varepsilon_{j}+U-\sigma+\delta_{j1}\bigl\langle n_{\bar{j}}\bigr\rangle U\lambda^{2}\tilde{K})}$
and $\sigma=\frac{\tilde{\Sigma}\bigl\langle n_{\bar{j}}\bigr\rangle U}{\varepsilon+\varepsilon_{j}+\tilde{\Sigma}-\delta_{j1}\lambda^{2}K}$.
Notice that we find

\begin{align}
\Sigma^{(j)}=\Sigma(1+\frac{U\bigl\langle n_{\bar{j}}\bigr\rangle}{\varepsilon-\varepsilon_{j}-U+i0^{+}})\label{eq:NSE_LEADS}
\end{align}
and

\begin{align}
\Sigma_{\text{{MBS1}}}^{\text{{full}}}=\Sigma_{\text{{MBS1}}}+\frac{U\bigl\langle n_{2}\bigr\rangle\Sigma_{\text{{MBS1}}}}{(\varepsilon-\varepsilon_{1}-U+i0^{+})}\nonumber \\
-U\bigl\langle n_{2}\bigr\rangle\lambda^{2}\tilde{K}_{U}(\frac{U\bigl\langle n_{2}\bigr\rangle}{\varepsilon-\varepsilon_{1}-U+i0^{+}}+1)\times\nonumber \\
(\Sigma_{\text{{MBS1}}}-\tilde{\Sigma}\lambda^{2}\tilde{K})\label{eq:MBS_full}
\end{align}
as the self-energies dressed by the interdots Coulomb repulsion. We
highlight that these expressions recover the well know results for
the noninteracting case $U=0$, as we found in Ref.\,\cite{key-16}.

\section{Results and Discussion}

\begin{figure}[!htb]
\begin{centering}
\includegraphics[clip,width=0.50\textwidth,height=0.30\textheight]{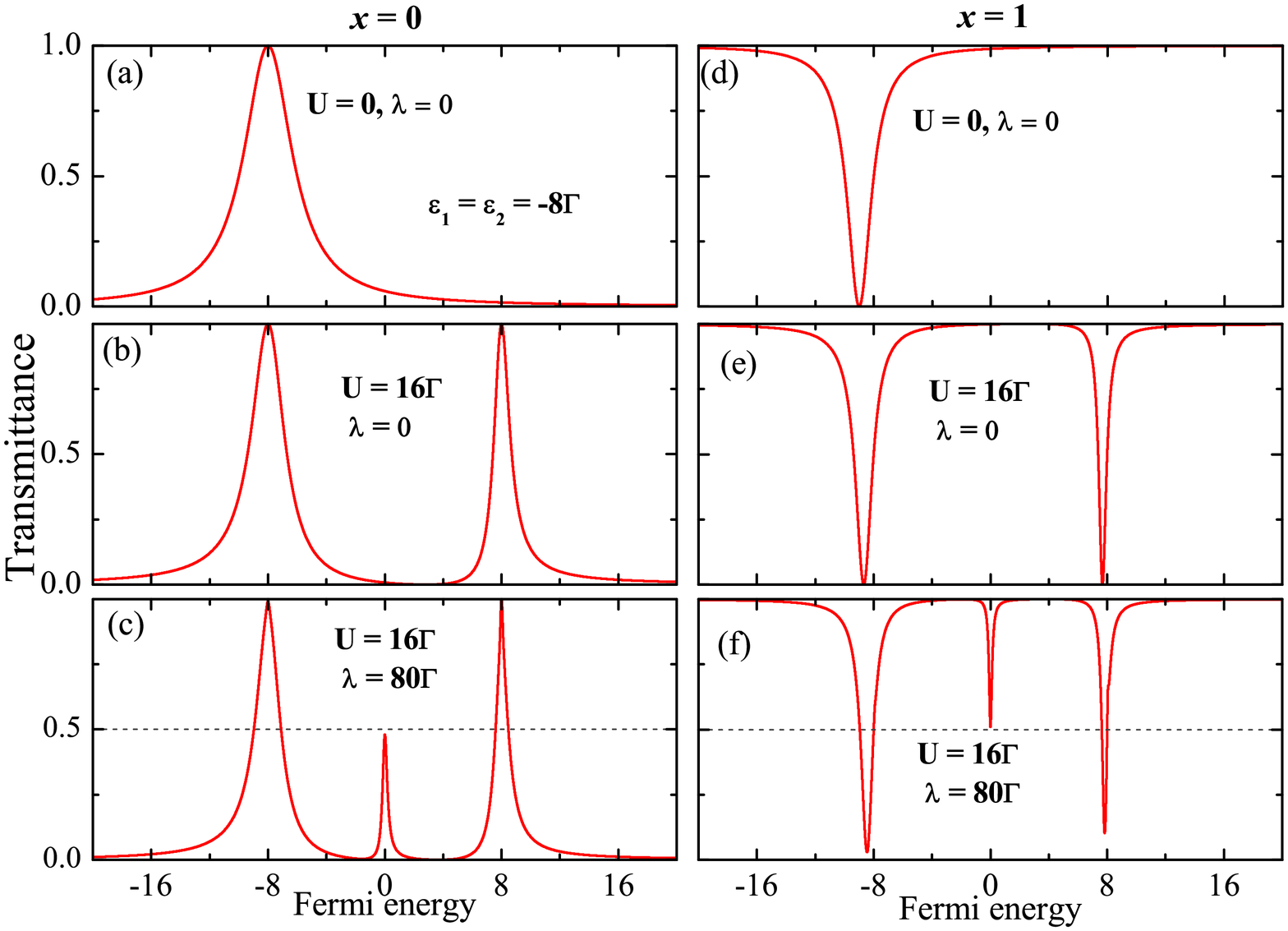}
\protect\protect\caption{\label{fig2}(Color online) Transmittance of Eq.\,(\ref{eq:TRANSMITTANCE})
as a function of the leads Fermi energy for two distinct Fano regimes.
The left panels {[}(a), (b) and (c){]} show the case wherein the electronic
tunneling occurs via QDs $(x=0$, $q_{b}\rightarrow\infty),$while
the right panels {[}(d), (e) and (f){]} exhibit the opposite Fano
regime, where the electronic path is preferentially through the metallic
leads $(x=1$, $q_{b}=0).$}

\par\end{centering}

\centering{}
\end{figure}

In what follows we discuss the effect of the Coulomb interdots correlation
in the Majorana signature. The entire analysis performed here is for
the case of a long wire, in such a way that the overlap between the
wave functions of the MBSs can be neglected $(\varepsilon_{M}=0)$. Additionally,
we adopt in our simulations the model parameters of the system Hamiltonian, in units of $\Gamma=2V^{2}\pi\rho_{0}$.
The transmittance profiles {[}Eq.\,(\ref{eq:TRANSMITTANCE}){]} as
a function of the Fermi energy for metallic leads are presented in
Fig.\,\ref{fig2}, in particular for the case of QDs in resonance
$(\varepsilon_{1}=\varepsilon_{2}=-8\Gamma)$. The left panels {[}Figs.\,\ref{fig2}(a), (b) and (c){]}
are for the Fano regime $q_{b}\rightarrow\infty$ $(x=0),$ where
the electronic path is completely through the QDs, while the right
panels {[}Figs.\,\ref{fig2}(d), (e) and (f){]} show the opposite
case, where the electrons travel preferentially via leads ($q_{b}=0$,
$x=1$). In Fig.\,\ref{fig2}(a), the Kitaev wire is decoupled from
the noninteracting QDs $(U=0$, $\lambda=0)$, and leads to a resonance
pinned at the energy levels of these QDs. For the interacting case
$(U=16\Gamma)$ {[}Fig.\,\ref{fig2}(b){]}, but still without wire,
the transmittance profile displays the Hubbard bands, represented
by two peaks pinned at $\varepsilon_{j}$ and $\varepsilon_{j}+U$,
respectively, where we consider the particle-hole symmetric point
$(2\varepsilon_{j}+U=0)$ of the system Hamiltonian. Fig.\,\ref{fig2}(c)
displays the case where the wire is strongly coupled to interferometer
with interacting QDs $(\lambda=80\Gamma$ and $U=16\Gamma)$. As one
can see, a zero-bias peak with 0.5 amplitude emerges making explicit
that the MBS 1 at the wire edge leaked into the QD 1. Such a panel
also exhibits two peaks fixed at $\varepsilon_{j}$ and $\varepsilon_{j}+U$
as in the upper case, but slightly narrower, which is an effect due
to the presence of the wire. The opposite Fano regime (right panels)
presents the same behavior showed in Figs.\,\ref{fig2}(a), (b) and (c),
but the peaks are replaced by dips, in agreement with the standard
Fano theory of interference \citep{key-29}.

\begin{figure}[!htb]
\begin{centering}
\includegraphics[width=0.5\textwidth,height=0.31\textheight]{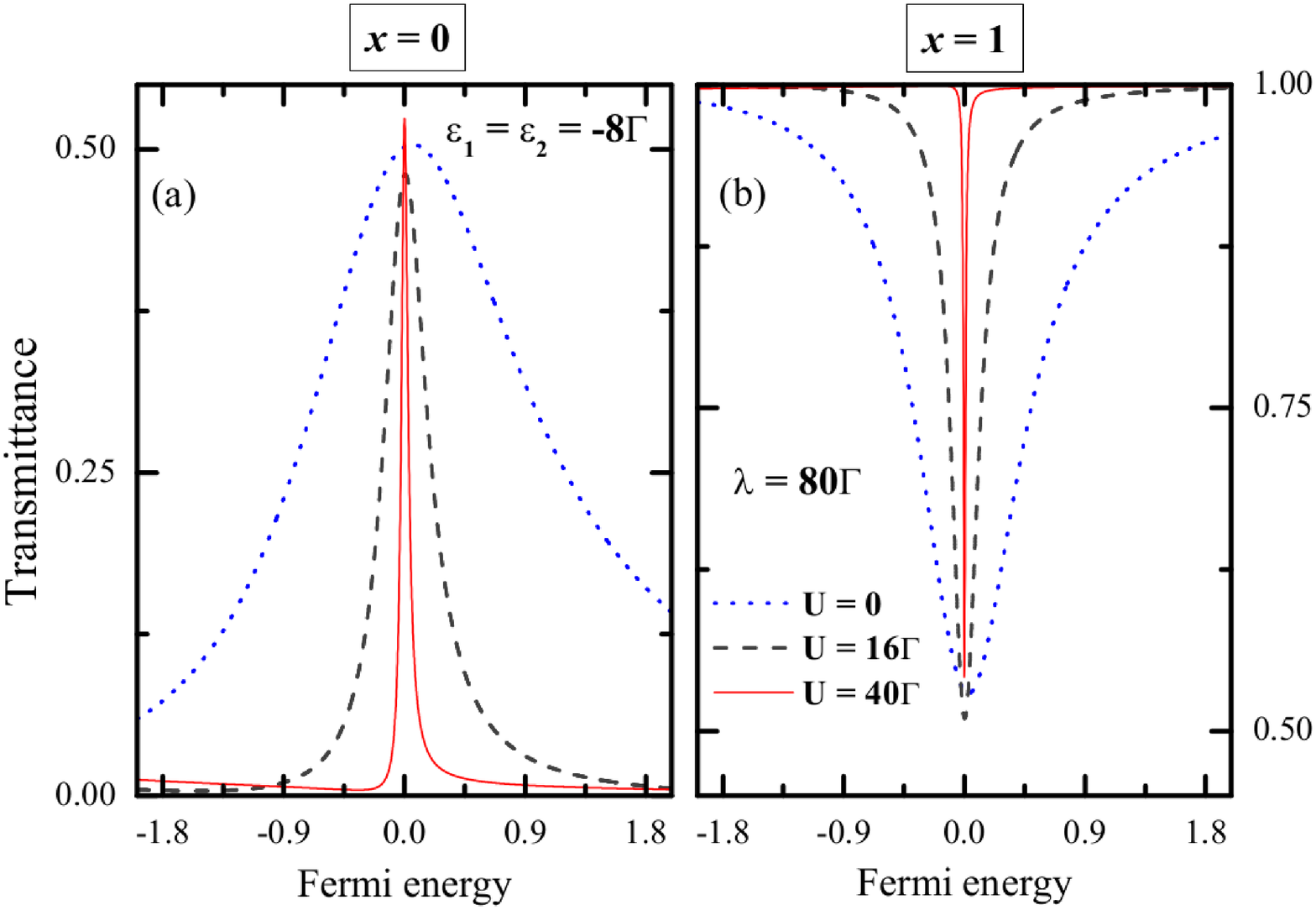}
\protect\protect\caption{\label{fig3}(Color online) Transmittance profiles of Eq.\,(\ref{eq:TRANSMITTANCE})
as a function of leads Fermi energy display the effect of Coulomb
interaction in the Majorana hallmark for distinct Fano regimes: (a)$x=0$,
$q_{b}\rightarrow\infty$ and (b)$x=1$, $q_{b}=0$.}

\par\end{centering}

\centering{}
\end{figure}

In Fig.\,\ref{fig3} we analyze the effect of the Coulomb repulsion
in the Majorana signature, namely the ZBA, for both Fano regimes.
The QDs are still in resonance and the wire is strongly coupled to
QD 1. The dotted-blue line shows the noninteracting case $(U=0)$,
where we can verify a ZBA characterized by a broad width in the two
Fano cases $(x=0$ and $x=1)$. Notice that there is a small difference
between the amplitudes: for the case where $x=0$ {[}Fig.\,\ref{fig3}(a){]}
the amplitude is exactly 0.5, while in $x=1$ {[}Fig.\,\ref{fig3}(b){]}
it is $<0.5$. Such a feature is due to the presence of the QD 2 and
a more detailed discussion can be found in Ref.\,\cite{key-15}. As
we increase the Coulomb interaction (dashed-black line for $U=16\Gamma$
and red line for $U=40\Gamma$), the ZBA becomes narrower and shows
a slight fluctuation around the 0.5 amplitude, verified in both Fano
regimes. This change in the ZBA width suggests that the Majorana state
lifetime within the QD 1 increases due to the electronic correlation
effects between the QDs, since this lifetime is inversely proportional
to such a width \citep{P. Phillips}.

\begin{figure}[!htb]
\begin{centering}
\includegraphics[width=0.5\textwidth,height=0.31\textheight]{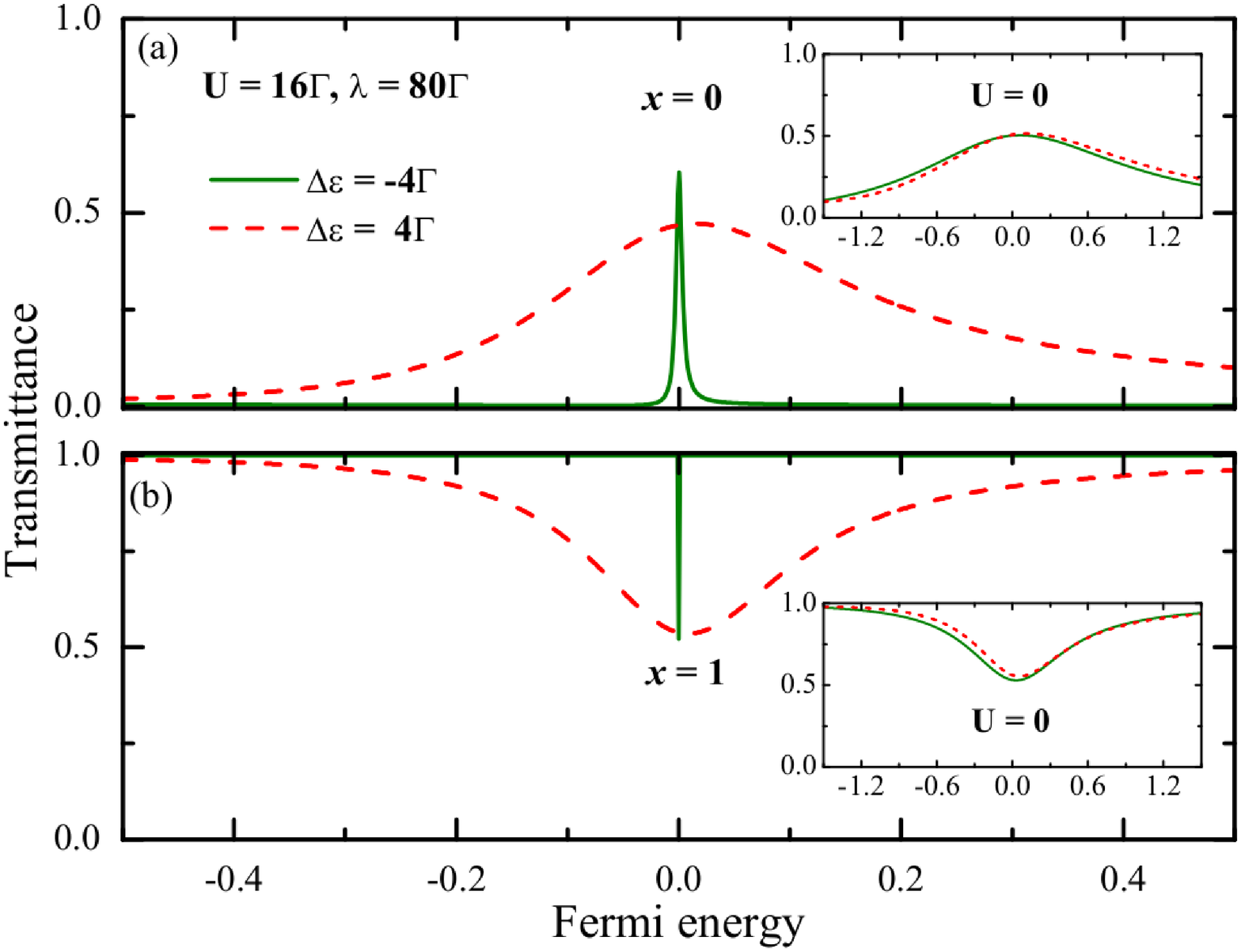}
\protect\protect\caption{\label{fig4}(Color online) Transmittance of Eq.\,(\ref{eq:TRANSMITTANCE})
as a function of leads Fermi energy, in the situation where the energy
levels of QDs are off-resonance, leading to a finite detuning $\Delta\varepsilon=\varepsilon_{2}-\varepsilon_{1}.$
The green line represents the case where $\varepsilon_{1}=-4\Gamma$
and $\varepsilon_{2}=-8\Gamma$, while the dashed-red line depicts
for $\varepsilon_{1}=-8\Gamma$ and $\varepsilon_{2}=-4\Gamma$. The
panels (a) and (b) represents the Fano regime $x=0$ $(q_{b}\rightarrow\infty)$
and $x=1$ $(q_{b}=0)$, where the corresponding insets describe the
noninteracting scenario. }

\par\end{centering}

\centering{}
\end{figure}

Fig.\,\ref{fig4} shows transmittance profiles {[}Eq.\,(\ref{eq:TRANSMITTANCE}){]}
as a function of leads Fermi energy for the interacting case, in the
situation where the energy levels of the QDs are off-resonance, leading
to a finite detuning $\Delta\varepsilon=\varepsilon_{2}-\varepsilon_{1}.$
The upper panel {[}Fig.\,\ref{fig4}(a){]} shows the behavior of the
transmittance for the Fano regime $x=0$. The Majorana peak is narrow
and has an amplitude slightly higher than 0.5 when $\varepsilon_{1}=-4\Gamma$
and $\varepsilon_{2}=-8\Gamma$ (green line). By making a swap in
the energy levels of the QDs, i.e. $\varepsilon_{1}=-8\Gamma$ and
$\varepsilon_{2}=-4\Gamma,$ the width of the Majorana peak increases
significantly and exhibits a 0.5 amplitude (dashed-red line). A similar
behavior is observed for the opposite Fano regime {[}Fig.\,\ref{fig4}(b){]},
but as in Figs.\,\ref{fig2} and \ref{fig3}, the peaks are replaced
by dips. Such fluctuations observed in the width and amplitude for
the ZBA arising from the MBS do not occur for the noninteracting case,
as can be seen in the insets of both panels. Thus, the tuning of the
energy levels for the QDs also plays an important role in the Majorana
state lifetime. This particular behavior suggests that experimentally,
the Majorana signature depends on the relative positions of the energy
levels for the QDs.

\section{Conclusions}

We have studied theoretically the effects of the Coulomb repulsion
in the Majorana signature in a system composed of a spinless double dot
interferometer with two interacting QDs, where one of them is strongly
coupled to the MBS hosted at one edge of a Kitaev wire within the
topological phase. We have found that the ZBA in the transmittance,
due to the MBS, appears even in the presence of interdots Coulomb
interaction. For QDs in resonance, the width of this Majorana hallmark
decreases as we increase the ratio between the strength of the Coulomb
repulsion and the dot-wire coupling $(U/\lambda)$, together with
a slight fluctuation around the $0.5$ amplitude. This narrowing of
the Majorana signature also occurs when a swap between the energy
levels for the QDs is performed. Such a narrowing is not verified
in the noninteracting case. These features then point out that the
ratio $U/\lambda$ and the relative position of the energy levels
(detuning) for the QDs, constitute the key ingredients ruling the
Majorana state lifetime in the QD next to the Kitaev wire. Our findings
shed new light into the ZBA of MBSs in the presence of Coulomb correlations
and can be helpful in the quest for Majorana signatures in condensed
matter systems.

\section*{Acknowledgments}

This work was supported by the Brazilian agencies CNPq, CAPES and
São Paulo Research Foundation (FAPESP) - Grant 2014/14143-0.

\end{document}